\documentclass[12pt]{article} 
\pdfoutput=1
\usepackage{mathrsfs}
\usepackage{amssymb}
\usepackage{verbatim}
\usepackage{graphicx,epsf,rotate} 
\usepackage{cite,color,xspace}
\usepackage{amsmath}
\begin{document}
\thispagestyle{empty}

\begin{center}
{\Large \bf Implications of diphoton searches for a \\Radion in the Bulk-Higgs Scenario} \\
\vspace{.2cm}
{\normalsize\sc
Peter Cox$^{a,}$\footnote{\texttt{pcox@physics.unimelb.edu.au}},
Anibal D.~Medina$^{b,}$\footnote{\texttt{anibal.medina@cea.fr}},\\
Tirtha Sankar Ray$^{c,}$\footnote{\texttt{tirthasankar.ray@gmail.com}}
 and Andrew Spray$^{d,}$\footnote{\texttt{a.spray.work@gmail.com}}}
\\[5mm]
{\small\textit{$^{a}$ARC Centre of Excellence for Particle Physics at the Terascale,\\
School of Physics, The University of Melbourne, Victoria 3010, Australia\\
$^{b}$ Institut de Physiqu\'e Th\'eorique, Universit\'e Paris Saclay, CNRS, CEA, F-91191 Gif-sur-Yvette, France\\
$^{c}$Department of Physics and Center for Theoretical Studies,\\ 
		   Indian Institute of Technology,
		   Kharagpur 721302, India \\
$^{d}$ Center for Theoretical Physics of the Universe, \\Institute for Basic Science (IBS), Daejeon, 34051, Korea}
}
\end{center}


\begin{abstract}
In this work we point out that the apparent diphoton excess initially presented by the ATLAS and CMS collaborations could have originated from a radion in the bulk Higgs scenario within a warped extra dimension. In this scenario the couplings of the radion to massive gauge bosons are suppressed, allowing it to evade existing searches. In the presence of mixing with the Higgs, due to the strong constraints from diboson searches, only points near what we denominate the alignment region were able to explain the diphoton signal and evade other experimental constraints. In light of the new measurements presented at ICHEP 2016 by both LHC collaborations, which do not confirm the initial diphoton excess,  we study  the current and future collider constraints on a radion within the bulk-Higgs scenario. We find that searches in the diphoton channel provide the most powerful probe of this scenario and already exclude large regions of parameter space, particularly for smaller warp factors. The radion has a sizeable branching ratio into top pairs and this channel may also give competitive constraints in the future. Finally, diHiggs searches can provide a complementary probe in the case of non-zero radion-Higgs mixing but strong alignment.

\end{abstract} 


~~~~{Keywords:} Higher-dimensional, Nambu-Goldstone bosons

~~~~{PACS number:} 11.10.Kk, 12.60.Fr, 14.80.Rt, 14.80.Va

\section{Introduction}

Not that long ago, the ATLAS and CMS collaborations reported a mild excess in the diphoton channel around an invariant mass of 750 GeV, with a few fb$^{-1}$ of data and collisions at $\sqrt{s}=13$~TeV~\cite{ATLAS1, ATLAS2, CMS1, CMS2}. It was uncertain whether this excess was compatible with a large or narrow width resonance (ATLAS preferred a somewhat broad resonance with $\Gamma\approx 45$~GeV while CMS obtained a better fit with a narrower width) and the events did not seem to contain additional energetic particles. Furthermore, the absence of a significant signal in the previous run of the LHC at $\sqrt{s}=8$~TeV may have favoured certain production modes. The excess inspired a number of theoretical investigations~\cite{Buttazzo:2015txu} - \cite{ Kaneta:2015qpf}.

In this work we advance an explanation for the apparent excess in the context of warped extra dimensional models via the production of a spin-0 excitation of the metric: the radion, which becomes massive once the metric is stabilized. Similarly, via the AdS/CFT correspondence the radion can be thought of as the pseudo-Nambu-Goldstone boson (pNGB) of the spontaneously broken conformal sector; the dilaton. Warped extra dimensions~\cite{Randall:1999ee} with the non-factorizable $AdS_5$ space are a popular setup to address the gauge hierarchy problem and are a practical tool to study strongly coupled conformal sectors via the duality language. In order to explain the diphoton excess we find it advantageous to consider the Higgs field responsible for electroweak symmetry breaking as a \emph{5D} scalar in the \emph{bulk} of the extra dimension. It was shown in Ref.~\cite{Cox:2013rva} that the coupling of the radion to the massive gauge bosons is then highly suppressed compared to the case of a brane-localized Higgs. This suppresses the radion decay width to $WW$ and $ZZ$, allowing a relatively light radion to avoid the stringent LHC constraints from diboson searches. In the 4D CFT language, a bulk Higgs corresponds to a generic composite scalar. Spontaneous symmetry breaking driven by a brane-localised potential corresponds to this scalar gaining a VEV only when the CFT runs to the IR. The fine-tuning of the electroweak sector is no worse than for a conventional brane-localised Higgs.

However, assuming the conventional RS value for the size of the extra dimension $kL \approx 35$, we find that the above scenario is unable to reproduce the observed signal strength without running afoul of $t\bar{t}$ constraints. This motivates us to consider theories with $kL < 35$, which enhance the radion coupling to diphotons. Models of this nature have been previously studied, most notably in the ``Little RS'' paradigm~\cite{Davoudiasl:2008hx}. The original Little RS implementations abandoned explaining the full Planck-electroweak hierarchy, in favour of explaining the flavour structure of the SM. They should be interpreted as calculable models for some 4D CFT with a UV cut-off $\mu_{UV} \ll M_{Pl}$ above which physics is no longer conformal. These models do not include 4D gravity, but it was later shown that the addition of brane-localised kinetic terms could remedy this without significantly modifying the phenomenology of the radion and other KK modes~\cite{George:2011sw}.

Unfortunately, the apparent excess initially observed in both experiments at the LHC quickly faded away with increased integrated luminosity, as was officially presented by both collaborations at ICHEP 2016, held in the city of Chicago, USA. Of course, Randall-Sundrum models provide a very well-motivated paradigm for physics beyond the Standard Model and continue to remain of significant interest irrespective of the diphoton excess. Moreover, from the theoretical point of view, a light radion/dilaton may show up in other beyond the Standard Model scenarios where a conformal sector is spontaneously broken with an additional small explicit breaking of the CFT. The radion is often the lightest new state in these models and can provide a rich phenomenology which is already being constrained by current searches at the LHC. In the particular case of a bulk Higgs, the radion phenomenology can differ significantly from the usual radion/dilaton models as pointed out in~\cite{Cox:2013rva}. Thus it is important to study the collider signatures of this scenario across the complete parameter space, beyond the regions originally motivated by the diphoton excess. 

In this paper we first briefly review the radion phenomenology in the bulk Higgs scenario, as well as relevant features of Little RS models. We demonstrate that limits from 8~TeV resonance searches and measurements of the 125~GeV Higgs already constrain the radion-Higgs mixing to be small, before presenting the region in parameter space that can be in consonance with the would-be diphoton excess in Section~\ref{sec:excess}. We also point out that such a state would have large branching ratios to ditop and digluon final states, possibly in addition to di-Higgs, which provide complementary probes to explore these scenarios. Due to the absence of a signal in the new LHC data, in Section~\ref{sec:afterICHEP} we study the current and future collider constraints on the radion in bulk-Higgs RS models, extending our analysis to encompass the full range of warp factors and radion masses. We find that currently the diphoton channel provides the most powerful probe of these kinds of scenarios and show the projected future reach of the LHC.

\section{The Model}\label{sec:model}

The 5D metric for a slice of $AdS_5$ with boundaries at $y=0$ (UV-brane) and $y=L$ (IR-brane) is given by,
\begin{equation}
ds^2=e^{-2ky} \eta_{\mu\nu}dx^{\mu}dx^{\nu}-dy^2\label{RS}\,,
\end{equation}
where $y$ is the extra spatial coordinate and $k$ is the curvature scale. The spin-0 fluctuations of the metric are given by,
\begin{equation} \label{RS-spin0}
ds^2=e^{-2ky-2F(x,y)} \eta_{\mu\nu}dx^{\mu}dx^{\nu}-(1+2F(x,y))^2dy^2\,.
\end{equation}
The extra dimension also needs to be stabilized, which can be accomplished by the introduction of a scalar sector that couples minimally to gravity and leads to a non-vanishing mass for the radion. In the small back-reaction limit, the physical radion field\footnote{Which is a linear combination of the radion and the scalar introduced to stabilized the extra dimension.} has an approximate profile of the form~\cite{DeWolfe:1999cp},
\begin{equation} \label{radionprofile}
F(x,y)\equiv e^{2(ky-kL)} \frac{1}{\Lambda_r}\hat{r}(x)\,,
\end{equation}
where $\Lambda_r$ sets the strength of the coupling to Standard Model fields. In the minimal RS model it is related to the 5D Planck mass, $M_5$, via $\Lambda_r=\sqrt{3M_5^3/k}\,e^{-kL}$. If we treat the RS framework simply as a tool to analyze a 4D CFT, then $M_5$ and hence $kL$ and $\Lambda_r$ are free parameters.

However, there is a cost to taking this approach, namely that the 4D graviton coupling strength in the usual RS model is set by $M_4 \approx \sqrt{M_5^3/k}$. If we take $\Lambda_r \sim$~TeV and $kL \ll 35$, we find that $M_4 \ll M_{Pl}$, which is trivially excluded by observations. There are several possible resolutions to this problem. Ref.~\cite{Davoudiasl:2008hx} imposed Dirichelet boundary conditions on the 5D gravity field at the UV brane, projecting out the graviton zero mode entirely. The resultant radion phenomenology was studied in~\cite{Davoudiasl:2010fb} and found to be unchanged. These theories do not reproduce 4D gravity, nor do they explain the full Planck-electroweak hierarchy. As such, they \emph{must} be interpreted as calculable models of a 4D CFT with a cut-off $M_5$ above which a UV completion is required. However, the physics of the composite states, in particular the Higgs and the radion (dilaton), will be protected from the details of this UV completion.

An alternative strategy was developed in Ref.~\cite{George:2011sw}, where the gravitational action is supplemented with a UV-brane-localised kinetic term\footnote{An IR-brane kinetic term can also be included. Demanding a physical radion kinetic term imposes $0 < v_L < 1$, and its affect on the graviton phenomenology is suppressed.}
\begin{equation}
 S_{UV} \supset \frac{v_0 M_5^3}{k} \int d^4x \, \sqrt{-g_0} R_0 \,,
\label{eq:UVterm}\end{equation}
where $v_0$ is a dimensionless parameter and $g_0$, $R_0$ are the brane-induced metric and curvature scalar respectively. This term suppresses the graviton normalisation, so that the 4D gravity scale is now
\begin{equation}
 M_4^2 = \frac{M_5^3}{k} \, \bigl(1 - e^{-2kL} + v_0 \bigr) \approx \frac{M_5^3}{k} \, v_0 \text{ for } v_0 \gg 1 \,.
\end{equation}
Sufficiently large values of $v_0$ then reproduce the correct 4D gravity scale. Ref.~\cite{George:2011sw} also showed in detail that the radion phenomenology is largely unchanged. This can be easily understood from the fact that the graviton is localised towards the UV brane, and is thus sensitive to a UV-localised term. In contrast, as shown in \eqref{radionprofile} the radion is localised towards the IR brane, and is thus insensitive to the term of Eq.~\eqref{eq:UVterm}.

From the point of view of the radion phenomenology, both approaches lead to the same conclusion: the radion profile, and hence its coupling strengths, are mostly unchanged. In what follows, we will make the simplifying assumption of using Eqs.~\eqref{RS-spin0} and \eqref{radionprofile} to describe the radion, since the precise details of the UV-localised physics are unimportant for our results.

The radion phenomenology has been extensively studied in the case that the Higgs scalar is localized on the IR brane. Here we will consider a bulk Higgs scenario given by the following 5D action
\begin{align} \label{eq:action}
 S & = \int_0^Ld^5x\,\sqrt{g}\left[\left(\frac{M_5^3}{2}+{\xi}H^{\dagger}H\right)R_5+{\vert}D_MH\vert^2-V(H)\right] +S_{brane} \,,
\end{align}
where the bulk radion-Higgs mixing is parameterized by $\xi$. We assume that the bulk and UV brane potentials are renormalizable and that the IR brane potential spontaneously breaks electroweak symmetry. Because the symmetry breaking takes place on the IR brane, the natural size of the 4D VEV is $\tilde{k} \equiv k e^{-k L} \sim \frac{1}{3} m_{KK}$, where $m_{KK}$ is the typical mass of the Kaluza-Klein (KK) resonanaces. This is the same as for a brane-localized Higgs with a tree-level potential, and implies a similar degree of fine-tuning, $v_{ew}^2/\tilde{k}^2 \sim \mathcal{O}(1)\%$, with $v_{ew} = 246$~GeV. Once we fine-tune the Higgs VEV, the Higgs mass is automatically of the same scale and no additional tuning is required.

It was shown in~\cite{Cox:2013rva} that once the Higgs is promoted to a 5D scalar field and detached from the IR brane into the bulk of the extra dimension, the radion phenomenology can change drastically. In particular, the radion couplings to the Higgs field and the massive gauge bosons are substantially modified compared to the more usual brane-Higgs scenario. Focusing on the coupling to massive SM gauge bosons and expanding the metric in its spin-0 fluctuations in the matter action we obtain
\begin{eqnarray}
\mathcal{S}_{matter}&=&\int d^4x dy\sqrt{g}\,(D_{\mu}H^{\dagger}D^{\mu}H)\nonumber\\
&=&\int d^4x dy\, e^{-4(A(y)+F(x,y))}(1+2F(x,y))e^{2(A(y)+F(x,y))}D_{\mu}H^{\dagger}D^{\mu}H\nonumber\\
&\approx &\int d^4x dy\, (1-4F^2(x,y)+\mathcal{O}(F(x,y)^3))e^{-2A(y)}D_{\mu}H^{\dagger}D^{\mu}H\,,
\end{eqnarray}
where in the last two lines indices are contracted using the Minkowski metric. We therefore see that the linear coupling of the radion to the gauge boson mass terms vanishes. Notice that this is a general result for the kinetic term of any scalar. It is a geometrical result from the 5D point of view, and holds even in pure $AdS_5$ space (without UV/IR branes)~\cite{Cox:2013rva}. We can also understand this result through the 4D dual picture, where the dilaton can only couple derivatively to conformally invariant operators. The gauge boson mass terms arise from the Higgs kinetic term, $D_{\mu}H D^{\mu}H$, where $H$ is the SM Higgs doublet. Lorentz invariance forbids a linear derivative radion coupling to this operator, and therefore at tree-level the radion only couples to the kinetic terms of the massive gauge bosons. We summarize the leading contributions to the other radion couplings in Table~\ref{couplingsgauge}.

\begin{table}[t]
\centering
\begin{tabular}{|c||c|}
\hline
\rule{0mm}{4mm}
 SM final state & Radion coupling\\[0.3em]
\hline
\rule{0mm}{5mm}
$f\bar{f}$ & $\frac{m_f}{\Lambda_r}$\\ [0.3em]
\hline
\rule{0mm}{5mm}
$WW$ & $-\frac{2}{\Lambda_r}\frac{1}{kL}$\\ [0.3em]
\hline
\rule{0mm}{5mm}
$ZZ$ & $-\frac{1}{\Lambda_r}\frac{1}{kL}$\\ [0.3em]
\hline
\rule{0mm}{5mm}
$hh$ & $\frac{1}{\Lambda_r}\left(2m^2_h-\frac{c_1}{2}m^2_r\right)$\\ [0.3em]
\hline
\rule{0mm}{5mm}
$\gamma\gamma$ & $-\frac{1}{\Lambda_r}\left(\frac{1}{kL}+\left[b_{QED}-\frac{4}{3}F_{1/2}(\tau_{t,r})\right]\frac{\alpha_{EM}}{2\pi}\right)$\\ [0.3em]
\hline
\rule{0mm}{5mm}
$gg$ & $-\frac{1}{\Lambda_r}\left(\frac{1}{kL}+\left[b_{QCD}-\frac{1}{2}F_{1/2}(\tau_{t,r})\right]\frac{\alpha_{3}}{2\pi}\right)$\\ [0.3em]
\hline
\end{tabular}
\caption{Couplings of the radion in the gauge basis adopted from Ref~\cite{Cox:2013rva}.}
\label{couplingsgauge}
\end{table}

As a pNGB, it's possible for the radion to be naturally lighter than the KK resonances~\cite{ Bellazzini:2013fga, Coradeschi:2013gda, Megias:2014iwa, Cox:2014zea}, which tend to lie in the few TeV range. Therefore, from a purely 4D effective field theory perspective, we can consider simply the SM with an additional scalar. The effective Lagrangian for the Higgs-radion fields, including a bulk mixing term $\xi R_5 H^{\dagger}H$ and the leading back-reaction contribution, is given by~\cite{Cox:2013rva}
\begin{align}\label{effact}
 { \mathcal{L}}_{eff} & = \frac{1}{2} \partial_{\mu} \hat{h}(x) \partial^{\mu} \hat{h}(x) 
 	- \frac{1}{2} m_h^2\hat{h}(x)^2 
	+ \frac{1}{2} \left(1+c_2 \frac{v_{ew}^2}{\Lambda_r^2} \right) 
		\partial_{\mu} \hat{r}(x)\partial^{\mu} \hat{r}(x) \notag \\ 
 & - \frac{1}{2} m_r^2 \hat{r}(x)^2 
 	- c_1 \frac{v_{ew}}{\Lambda_r} \partial_{\mu} \hat{h}(x) \partial^{\mu} \hat{r}(x) 
	- c_3 \frac{v_{ew}}{\Lambda_r} m_r^2 \hat{h}(x) \hat{r}(x)\,,
\end{align}
where $c_1$, $c_2$ and $c_3$ are $\mathcal{O}(1)$ numerical coefficients and $\hat{h}(x)$ is the Higgs scalar. The $c_i$ can in principle be given in terms of wavefunction overlap integrals, but are model-dependent on the details of the radion stabilization.

As is clear from Eq.~\eqref{effact}, there is kinetic and mass mixing between the radion and the Higgs. One can diagonalize the kinetic term by going to a new basis 
\begin{equation}
 \hat{h} = h' + c_1\frac{v_{ew}}{\Lambda_r}\, \frac{r'}{Z}\,, \qquad \hat{r}=\frac{r'}{Z}\,,
\end{equation}
where $Z^2=1+(c_2+c_1^2)v_{ew}^2/\Lambda_r^2$. This transformation decouples the kinetic mixing but introduces additional mass mixing terms. The mass matrix in the basis $(r',h')$ then takes the form:
\begin{eqnarray}
M  
=
\left(
\begin{array}{ccc}
 \frac{m_r^2}{Z^2}+\frac{1}{Z^2}\frac{v_{ew}^2}{\Lambda_r^2}(c_1^2 m_h^2+2c_1 c_3 m_r^2) &  \frac{1}{Z}\frac{v_{ew}}{\Lambda_r}(c_1 m_h^2+c_3m_r^2)\\
 \frac{1}{Z}\frac{v_{ew}}{\Lambda_r}(c_1 m_h^2+c_3m_r^2) & m_h^2
\end{array}
\right)~.
\end{eqnarray}
An additional unitary transformation brings us to the mass basis $(r, h)$, where $r$ ($h$) is the 750~GeV radion-like (125~GeV Higgs-like) field. We can define the relation between the gauge and mass bases as
\begin{equation}\label{eq:mtogtrans}
 \hat{r}(x) = c_{rr} \, r(x) + c_{rh} \, h(x) \,, \qquad \hat{h}(x) = c_{hr} \, r(x) + c_{hh} \, h(x) \,.
\end{equation}
Note that because of the non-unitary transformation, $c_{ih}^2 + c_{ir}^2 \neq 1$ for $i = h, r$. However, this definition is useful to determine the couplings in the mass basis. In fact, notice that $c_{hr}$ and $c_{rh}$ provide respectively the SM Higgs-like and radion-like couplings of the heavier and lighter mass eigenstates $r(x)$ and $h(x)$, and moreover due to the non-unitarity transformation among the different basis, $c_{hr}=0$ does not imply that $c_{rh}=0$ and vice versa.

\section{The Diphoton Excess: pre-ICHEP 2016} \label{sec:excess}

We use the best-fit cross sections for the diphoton excess from the CMS combined 8+13~TeV analysis~\cite{CMSupdate} and extracted from the ATLAS analysis~\cite{ATLAS1, ATLAS2} in Ref.~\cite{Buttazzo:2015txu}:
\begin{align}
 \mu^{\text{ATLAS}}_{13~\text{TeV}} & = \sigma(pp \to
 S)_{13~\text{TeV}} \times \mathcal{B}(S \to \gamma\gamma) =
 10^{+4}_{-3} \text{ fb,} \label{atlas} \\ \mu^{\text{CMS}}_{13~\text{TeV}} &
 = \sigma(pp \to S)_{13~\text{TeV}} \times \mathcal{B}(S \to
 \gamma\gamma) = 3.7^{+1.5}_{-1.3} \text{ fb.} \label{cms}
\end{align}
We also summarise in Table~\ref{tab:xsecs} the most important constraints from 8~TeV resonance searches. The radion couplings to light fermions are suppressed by the fermion mass and so set no meaningful constraints.

\begin{table}
 \centering
 \begin{tabular}{|c|c|}
  \hline
  Final State & Observed Bound \\
  \hline
  $t\bar{t}$ & $<$ 300~fb~\cite{Chatrchyan:2013lca} \\
  $WW$ &$<$ 38~fb~\cite{Khachatryan:2015cwa,Aad:2015agg} \\
  $ZZ$ & $<$ 12~fb~\cite{Aad:2015kna} \\
  $Z\gamma$ & $<$ 4.0~fb~\cite{Aad:2014fha,CMS:2016all} \\
  $\gamma\gamma$ & $<$ 1.4~fb~\cite{Aad:2015mna,Khachatryan:2015qba} \\
  $hh$ & $<$ 36~fb~\cite{Aad:2015xja} \\
  $jj$ & $<$ 2.5~pb~\cite{Aad:2014aqa,CMS:2015neg}\\
  \hline
 \end{tabular}
 \caption{Constraints on the radion from 8~TeV resonance searches.}\label{tab:xsecs}
\end{table}

From Table~\ref{tab:xsecs}, we see that the constraints on a putative 750~GeV resonance decaying to gauge bosons are an order of magnitude stronger than on decays to $t\bar{t}$. In contrast, a scalar of that mass with Higgs-like couplings will dominantly decay to $WW$ and $ZZ$. This has important consequences for our model since the radion and the Higgs generically have both mass and kinetic mixing, as shown in Eq.~\eqref{effact}. In order to sufficiently suppress the heavier state decay to gauge bosons, it must be radion-like with any Higgs-like couplings strongly suppressed. Large mixing between the two states would also modify the properties of the 125 GeV scalar resonance discovered at the LHC. Decays of the heavier state to $hh$ could also affect the Higgs measurements, however this contribution will turn out to be negligible in the viable regions of parameter space.

\begin{figure}[h]
 \centering
 \includegraphics[height=7cm]{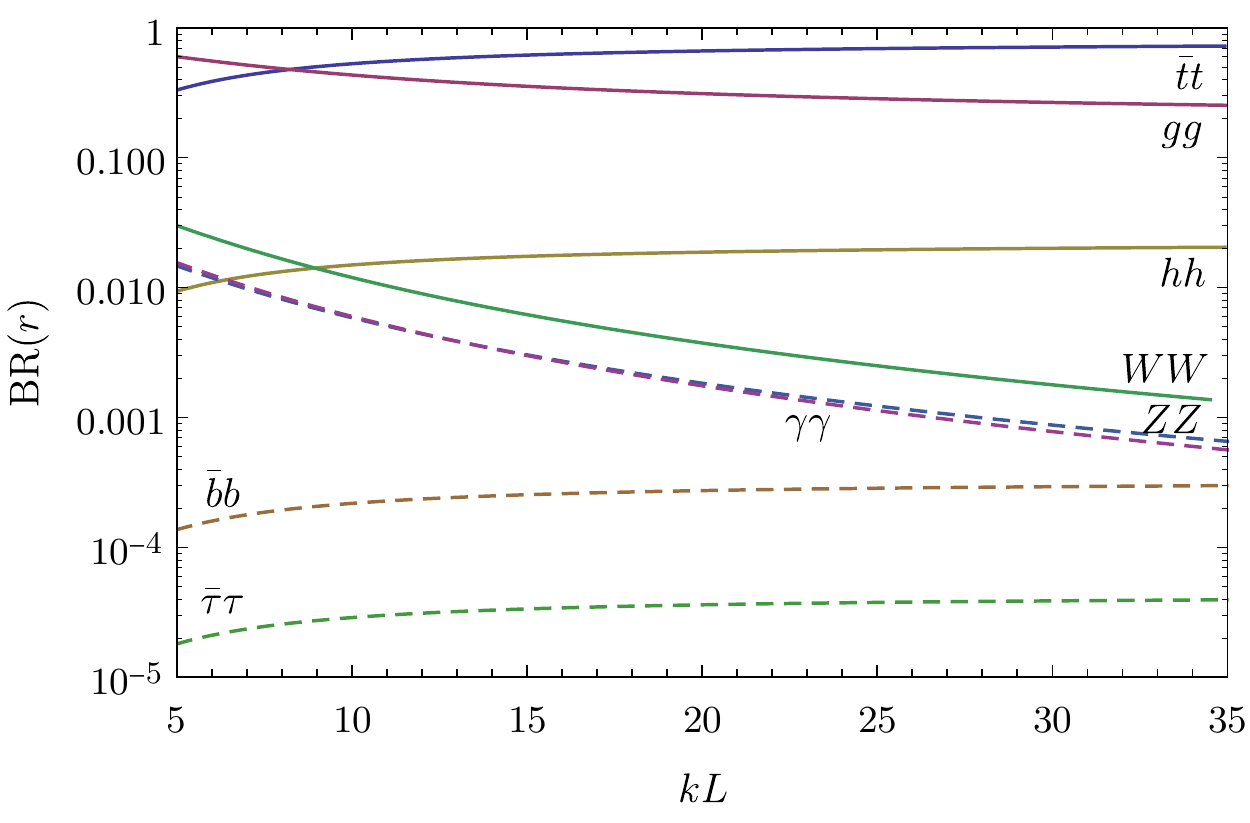}
 \caption{\small Branching ratios for a 750 GeV radion to various SM final states. This is for the bulk Higgs setup in the zero mixing limit with the SM Higgs.}
 \label{fig:branching}
\end{figure}

\begin{figure}[h]
 \centering
 \includegraphics[height=7cm]{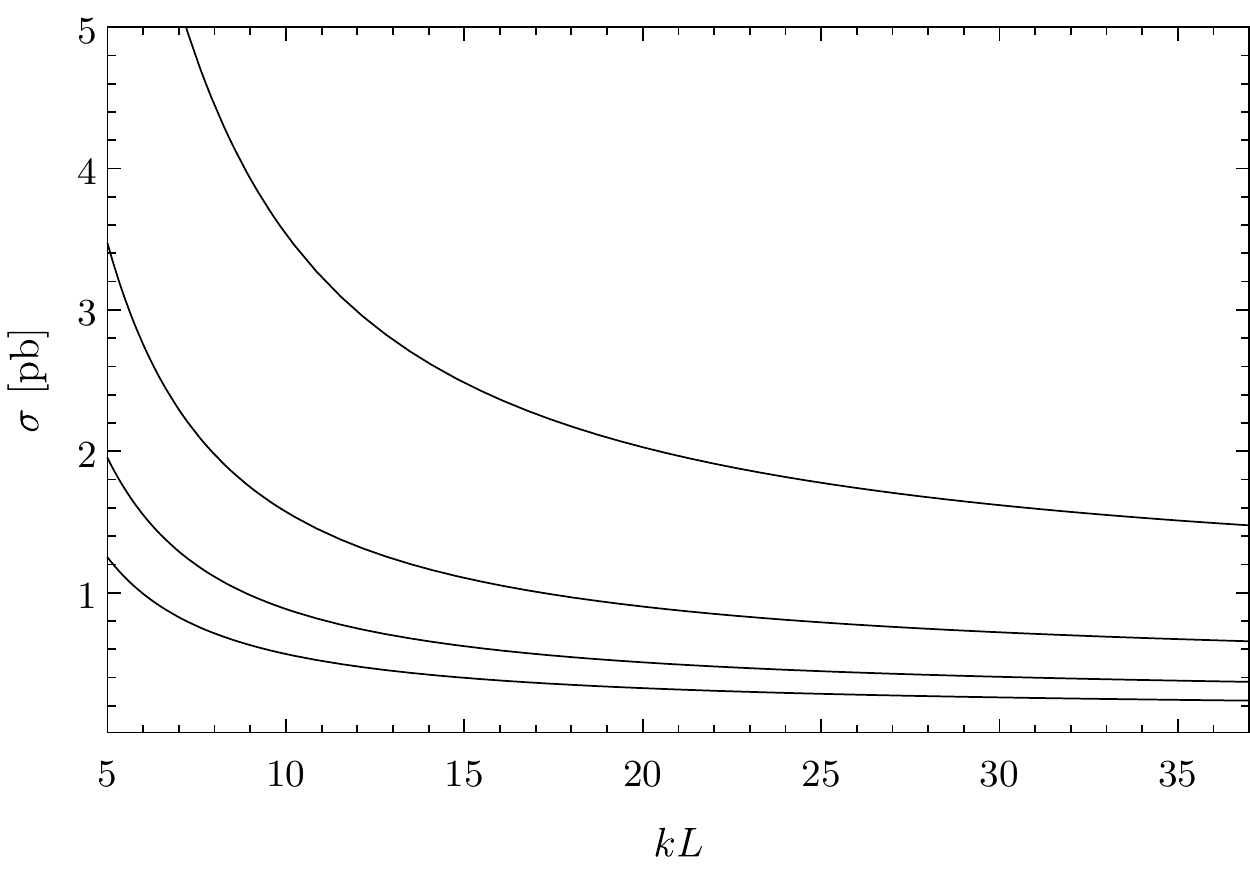}
 \caption{\small Production cross section for a 750 GeV radion via gluon-gluon fusion. From top to bottom the curves correspond to $\Lambda_r=2,\,3,\,4,\,5\,$TeV in the limit of zero mixing with the SM Higgs.}
 \label{fig:xsec}
\end{figure}

With the radion couplings to massive gauge bosons suppressed, inspecting Table~\ref{couplingsgauge} reveals that the other potentially large couplings are to gluons (due to the QCD anomaly), to top-quark pairs (since they are the heaviest SM fermion), and to the lighter Higgs-like state. The decay $r \to t\bar{t}$ is enhanced over $r \to hh$ by colour and spin degrees of freedom, enough that the former typically sets the dominant limits despite the stronger constraints on $hh$ final states listed in Table~\ref{tab:xsecs}. Note from Table~\ref{couplingsgauge} that all radion couplings scale with $\Lambda_r$ in the same way. Therefore in the zero-mixing limit the branching ratios depend only on the size of the extra dimension $kL$. We plot those in Figure~\ref{fig:branching}, and see that $\mathcal{B} (r \to t\bar{t}) \approx 40 \,\mathcal{B} (r \to hh)$.

Our model contains only four free parameters: $\Lambda_r$, $kL$, $c_1$ and $c_3$. The observed masses for the Higgs and the diphoton excess fix $m_r$, $m_h$ in terms of these parameters, while $c_2$ has a negligible effect since it comes with an additional $v_{ew}/\Lambda_r$ supression factor. In order to identify the regions of parameter space which both satisfy the current experimental bounds and are consistent with the observed diphoton excess we peform a scan over these parameters. We take flat priors in the range $-2\leq c_1,c_3\leq2$, $5\leq kL\leq35$ and $1\leq\Lambda_r\leq5\,$TeV and require that the 750~GeV state satisfy the constraints from the resonance searches in Table~\ref{tab:xsecs}, while the 125~GeV state is consistent with the measured Higgs signal strengths~\cite{Higgscomb} at the 2-sigma level. The radion production cross sections are obtained by scaling the Higgs 8~TeV cross section~\cite{Heinemeyer:2013tqa} by the radion effective coupling to gluons and, for $\sqrt{s}=13$~TeV, by the parton luminosity ratio:
\begin{equation}
 \frac{\sigma (pp \to r)_{13~\text{TeV}}}{\sigma (pp \to r)_{8~\text{TeV}}} \approx 4.7 \,.
\end{equation}
The resulting cross section is shown in Figure~\ref{fig:xsec}.

\begin{figure}[t]
 \begin{minipage}[t]{0.5\textwidth}
  \centering
  \includegraphics[width=\textwidth]{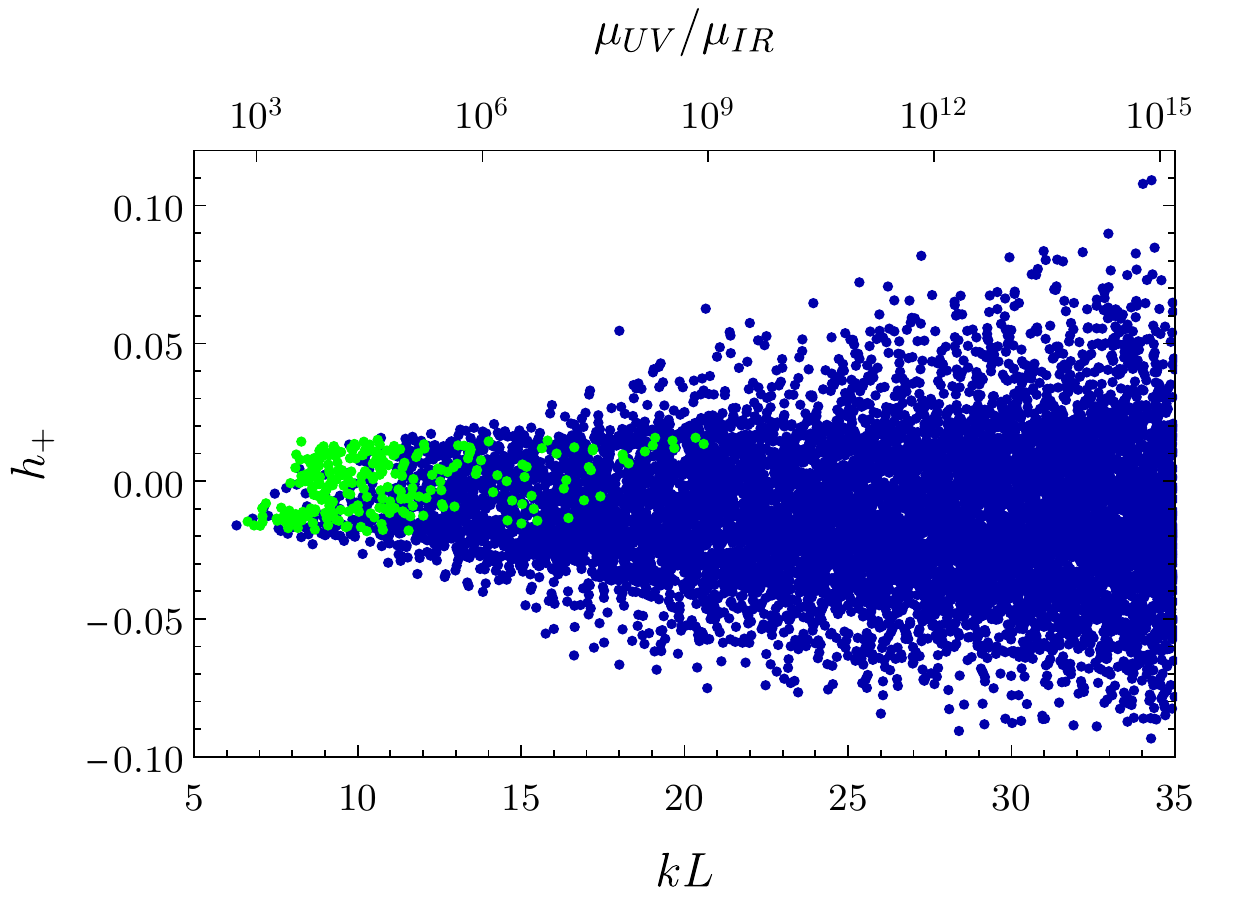}
 \end{minipage}
 \begin{minipage}[b]{0.5\textwidth}
  \centering
  \includegraphics[width=0.9\textwidth]{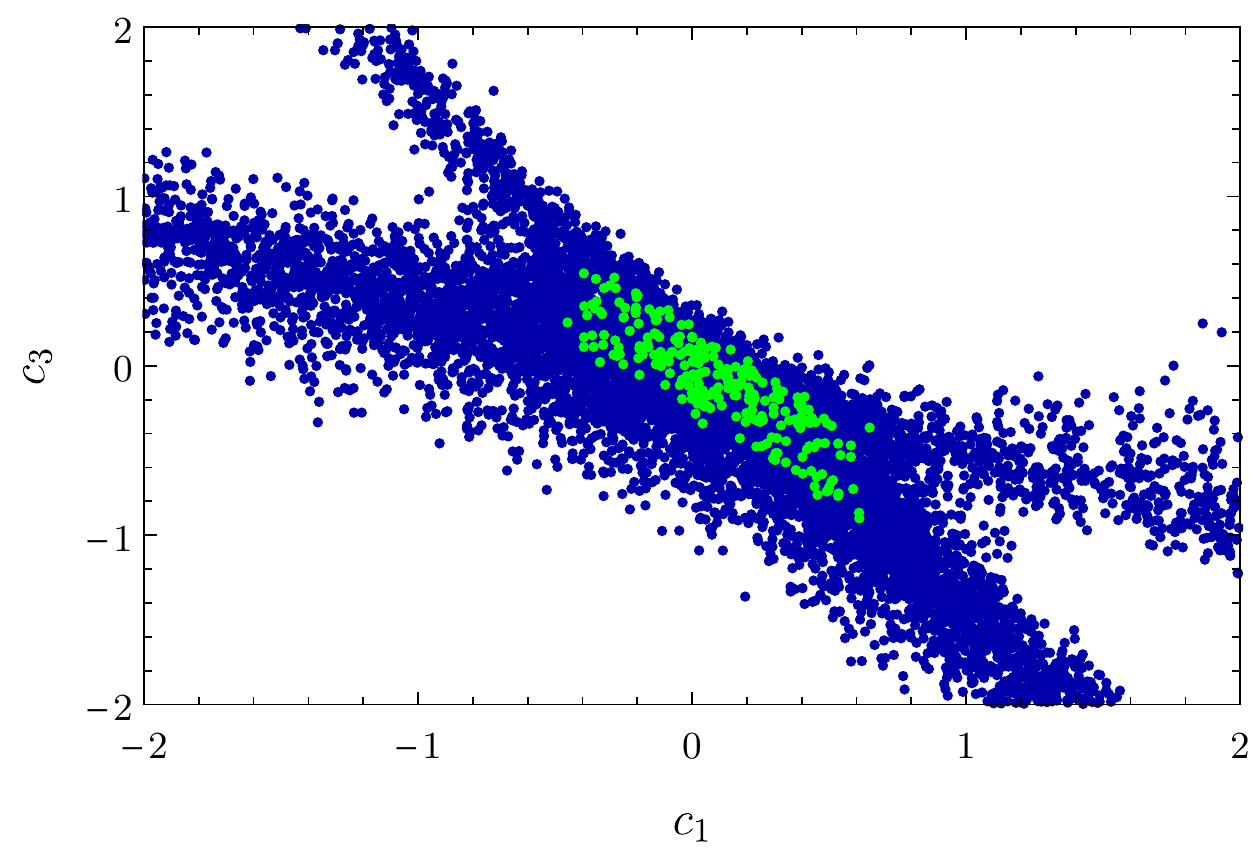}
 \end{minipage}
 \caption{\small Results of the ${\Lambda_r,\,kL,\,c_1,\,c_3}$ parameter scan. All points shown satisfy the constraints from resonance searches and Higgs measurements, while the green points are additionally consistent with the diphoton excess. We have taken the cross section for the diphoton excess to be the weighted average of Eqs.~\eqref{atlas} and \eqref{cms}, giving $4.9\pm1.3\,$fb.}
 \label{fig:scan}
\end{figure}

The results of the scan are shown in Figure~\ref{fig:scan}, where all points satisfy the current experimental constraints and the green points are also consistent with the observed diphoton excess. In the left panel we have plotted the 750~GeV component of the Higgs gauge state, $c_{hr}$, as a function of $kL$, clearly showing that the current data already enforces small Higgs-radion mixing. Remember that $c_{hr}$ gives a measure of how much Higgs-like couplings the heavier eigenstate possesses, and that for $c_{hr}\approx 0$ the heavier eigenstate has vanishing Higgs-like couplings. Large mixing is immediately excluded by measurements of the 125~GeV state; however it turns out that for much of the parameter space resonance searches in the $ZZ$ and $WW$ final states provide the stronger constraint. The bound on the mixing is therefore weaker at large $kL$ and large $\Lambda_r$, where the radion production cross section is reduced. Furthermore, the points consistent with the diphoton excess (green points) are concentrated at small $kL$ where the radion branching ratio to photons is enhanced and moreover the mixing is more strongly constrained from the massive diboson searches.

In the $v_{ew}/\Lambda_r\ll1$ limit, $c_{hr}$ is given in terms of the Lagrangian parameters by
\begin{equation}
 c_{hr}=(c_1+c_3)\,\frac{m_r^2}{m_r^2-m_h^2}\frac{v_{ew}}{\Lambda_r}+\mathcal{O}\left(\left(v_{ew}/\Lambda_r\right)^3\right)\,,
\end{equation}
and hence the heavier state is strongly radion-like along the line\footnote{The lighter state is not necessarily strongly Higgs-like for $c_1=-c_3$ due to the non-unitary transformation. That is instead determined by $|c_{rh}|=\frac{c_1m_h^2+c_3m_r^2}{m_h^2-m_r^2}\frac{v_{ew}}{\Lambda_r}+\mathcal{O}((v_{ew}/\Lambda_r)^3)\,$, where $c_{rh}=0$ corresponds to a purely Higgs-like state.} $c_1=-c_3$, which we refer to as the alignment case. This is evident in the right panel of Figure~\ref{fig:scan} where the allowed points are all located along the diagonal. Notice that there are no allowed points with $c_1\approx-c_3$ at large values of $c_1$. This region of parameter space is excluded by $hh$ resonance searches due to the enhanced radion coupling to $hh$ (see Table~\ref{couplingsgauge}). Enhancing the coupling to $hh$ also results in a decreased branching ratio to photons, hence the points satisfying the diphoton excess are concentrated around small values of $c_1$. These points are also aligned along the diagonal; the off-diagonal points corresponding to increased mixing generally have larger values of $kL$, again suppressing the branching ratio to photons.

\begin{figure}[h]
 \begin{minipage}[t]{0.5\textwidth}
  \centering
  \includegraphics[width=\textwidth]{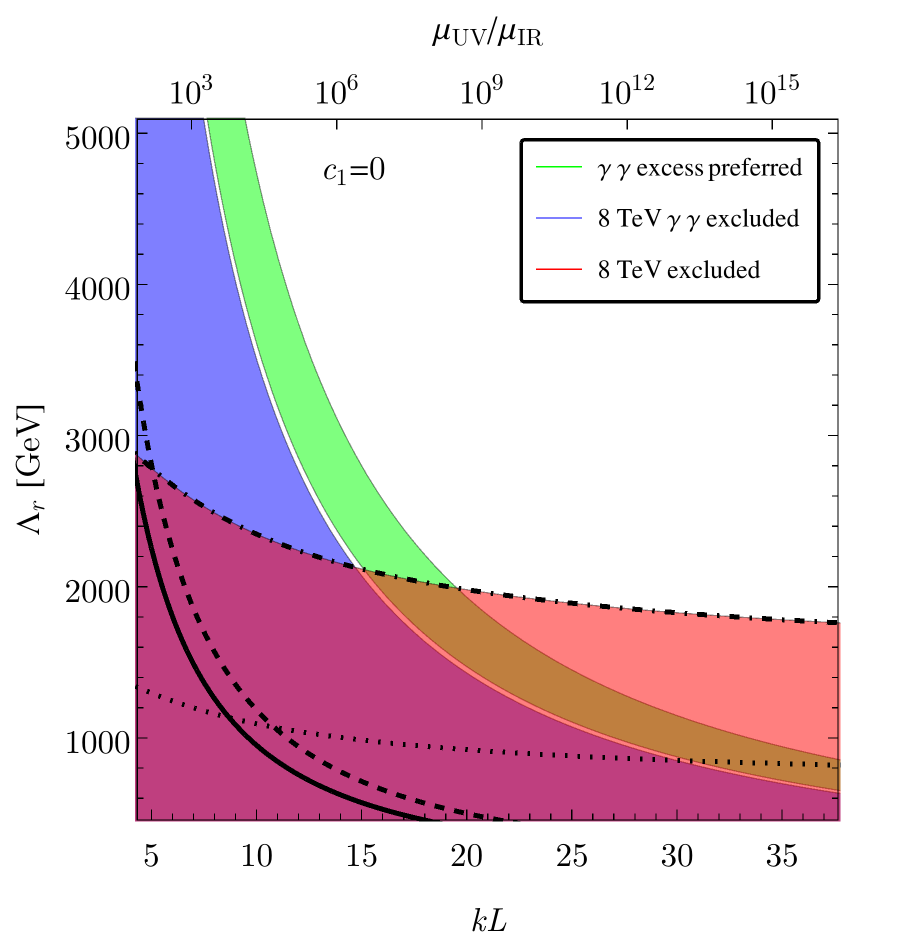}
 \end{minipage}
 \begin{minipage}[b]{0.5\textwidth}
  \centering
  \includegraphics[width=\textwidth]{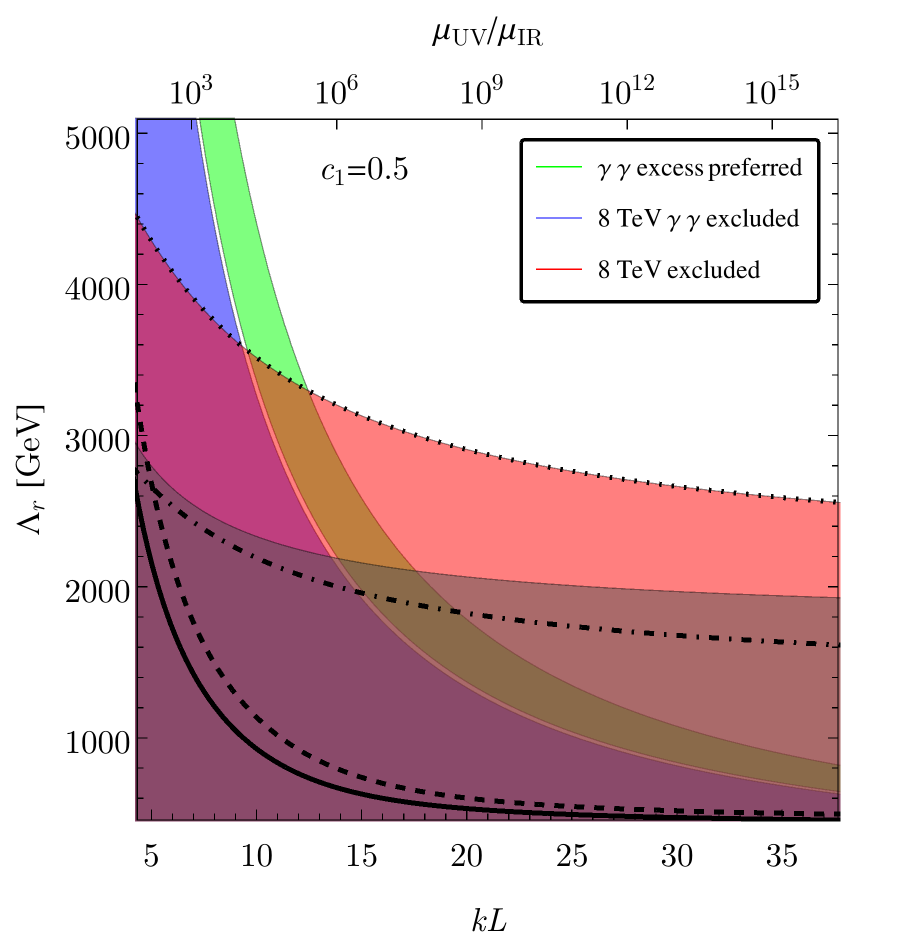}
 \end{minipage}
 \caption{\small The region of parameter space consistent with the observed diphoton excess is shaded in green, while the blue and red regions are excluded by the 8~TeV diphoton and other resonance searches respectively, and the grey region is excluded by Higgs measurements. Regions below and to the left of the solid, dashed, dotted and dot-dashed lines are individually excluded by the $WW$, $ZZ$, $hh$ and $tt$ searches listed in Table~\ref{tab:xsecs}. The left (right) panels are for $c_1=c_3=0$ ($c_1=-c_3=0.5$).}
 \label{fig:diphoton}
\end{figure}

It is clear that current experimental constraints impose small Higgs-radion mixing. Providing an explanation for the diphoton excess additionally restricts us to the region where $|c_1|\lesssim0.5$ and requires that we be quite close to the alignment case. Hence we now restrict ourselves to the case $c_1=-c_3$ in order to explore the ($kL$, $\Lambda_r$) parameter space in detail. In Figure~\ref{fig:diphoton} we show the regions in the $kL$--$\Lambda_r$ plane consistent with the observed di-photon excess while remaining unconstrained from other LHC searches. We have taken two benchmark points corresponding to $c_1=0$ and $c_1=0.5$.

Focusing on the left panel of Figure~\ref{fig:diphoton} ($c_1=0$), we see that $t\bar{t}$ searches give the approximate bound $\Lambda_r \gtrsim 2$~TeV, while obtaining a sufficiently large diphoton signal requires taking smaller values of $kL \lesssim 20$ than those used to obtain the Planck-electroweak hierarchy in RS-models ($kL\sim 35$). This can be understood by noting that to simultaneously have a 6~fb diphoton signal at 13~TeV while avoiding the constraints from ditop searches, we must have
\begin{equation}
 \frac{\mathcal{B} (r \to t\bar{t})}{\mathcal{B} (r \to \gamma\gamma)} < \frac{300}{6} \times \frac{\sigma (pp \to r)_{13~\text{TeV}}}{\sigma (pp \to r)_{8~\text{TeV}}} \approx 235\,.
\end{equation}
As can be seen from Figure~\ref{fig:branching}, this is not satisfied at $kL = 35$. Decreasing the size of the extra dimension increases the coupling to $\gamma\gamma$ (and $gg$, increasing the production cross section) allowing us to satisfy this constraint. In contrast, attempting to tune the mass and kinetic mixings between the radion and the Higgs to suppress the decay $r \to t\bar{t}$ while maintaining a sufficiently large production cross section reintroduces constraints from either the decays to $WW$ or $hh$, as can be inferred from Figure~\ref{fig:scan}. 

Moving away from $c_1=0$ leads to an increased branching ratio to $hh$ and this channel provides the strongest constraint for $c_1\gtrsim0.35$, as can be seen in the right panel of Figure~\ref{fig:diphoton}. Satisfying this constraint in addition to the diphoton excess then requires even smaller values of $kL$. Also note that while the heavier state is radion-like when $c_1=-c_3$, the lighter state is not necessarily Higgs-like due to the non-unitary transformation arising from the kinetic mixing. This leads to bounds from measurements of the Higgs couplings, which exclude the region $\Lambda_r\lesssim2\,$TeV. However the strongest constraint still comes from the diHiggs searches, which in the particular case of $c_1=0.5$ push the scale $\Lambda_r \gtrsim 3.5\,$~TeV.

Let us briefly comment on the theoretical implications of our preferred values for $kL$. Once we have resolved the graviton issue as discussed in Section~\ref{sec:model}, the primary constraint on $\mu_{UV} = \mu_{IR} \, e^{kL} \sim \Lambda_r \, e^{kL}$ comes from higher-dimensional operators generated at that scale. In particular, Ref.~\cite{Davoudiasl:2008hx} showed that the bounds from flavour operators demand that $\mu_{UV} \gtrsim 10^3$~TeV, which is satisfied for all points we show in Figure~\ref{fig:diphoton}.

Finally we address future prospects for testing this model. From Figure~\ref{fig:branching}, we see that the dominant decay mode of the radion is to $t\bar{t}$ and $gg$. While it might be difficult to observe the di-jet final state at early runs, the possibility to explore this scenario with the di-top final state is certainly within the reach of the LHC run 2. When mixing is allowed it is clear that searches involving massive gauge boson final states will be the main tool in constraining/discovering these kinds of models. However in the particular alignment case $c_1\approx -c_3$, where the heavier eigenstate is strongly radion-like, an interesting situation may arise in which, due to the linear $c_1$ dependence of the radion-diHiggs coupling, searches in the di-Higgs final state become the best way to probe the model for $c_1\gtrsim 0.35$. We should make clear that both the diboson and the di-Higgs searches are more model dependent whereas the di-top signature is relatively insensitive to assumptions about the mixing and thus a particularly promising channel. Additionally, we note that the width of the radion is quite small, $\Gamma \lesssim 1$~GeV. If ATLAS is correct that the excess is generated by a broad resonance, that would disfavour this model as a possible explanation. However it is too soon yet to say anything concrete regarding the width and CMS seems to slightly favour a narrower resonance. Lastly, we have worked in the 4D effective theory of Eq.~\eqref{effact}, but in the full theory we expect additional resonances at masses of a few TeV. If these are not too heavy, they might also be discoverable.

\section{The status of the model after ICHEP 2016} \label{sec:afterICHEP}

The non-observation of any significant excess in the larger $15\,\text{fb}^{-1}$ dataset from Run-II of the LHC at $\sqrt{s}=\,13\,$TeV, as presented at ICHEP 2016, significantly reduces the local significance of previous hints of a possible diphoton resonance at 750~GeV \cite{ATLAS1, ATLAS2, CMS1, CMS2}. This new data therefore excludes the possibility of a radion with the mass and couplings discussed in the previous section. However, the recently released results can instead be used provide the most up-to-date constraints on the radion phenomenology within the bulk Higgs framework, providing a significant improvement on the previous constraints. In this section we determine the constraints on the model parameter space and present the future prospects for the LHC to explore this scenario. 

We shall continue to focus on the alignment limit ($c_1$=$-c_3$). There are two reasons to consider this limit: (i) as we found in the previous section for $m_r=750\,$GeV, existing constraints from measurements of the Higgs couplings and direct searches in the $WW/ZZ$ final states are already pushing us into this region of parameter space; (ii) in this limit the radion phenomenology in the bulk Higgs scenario differs most significantly from the case of a brane-localised Higgs, which is the most standard and well-studied warped radion scenario.

\begin{figure}[t]
 \begin{minipage}[t]{0.5\textwidth}
  \centering
  \includegraphics[width=\textwidth]{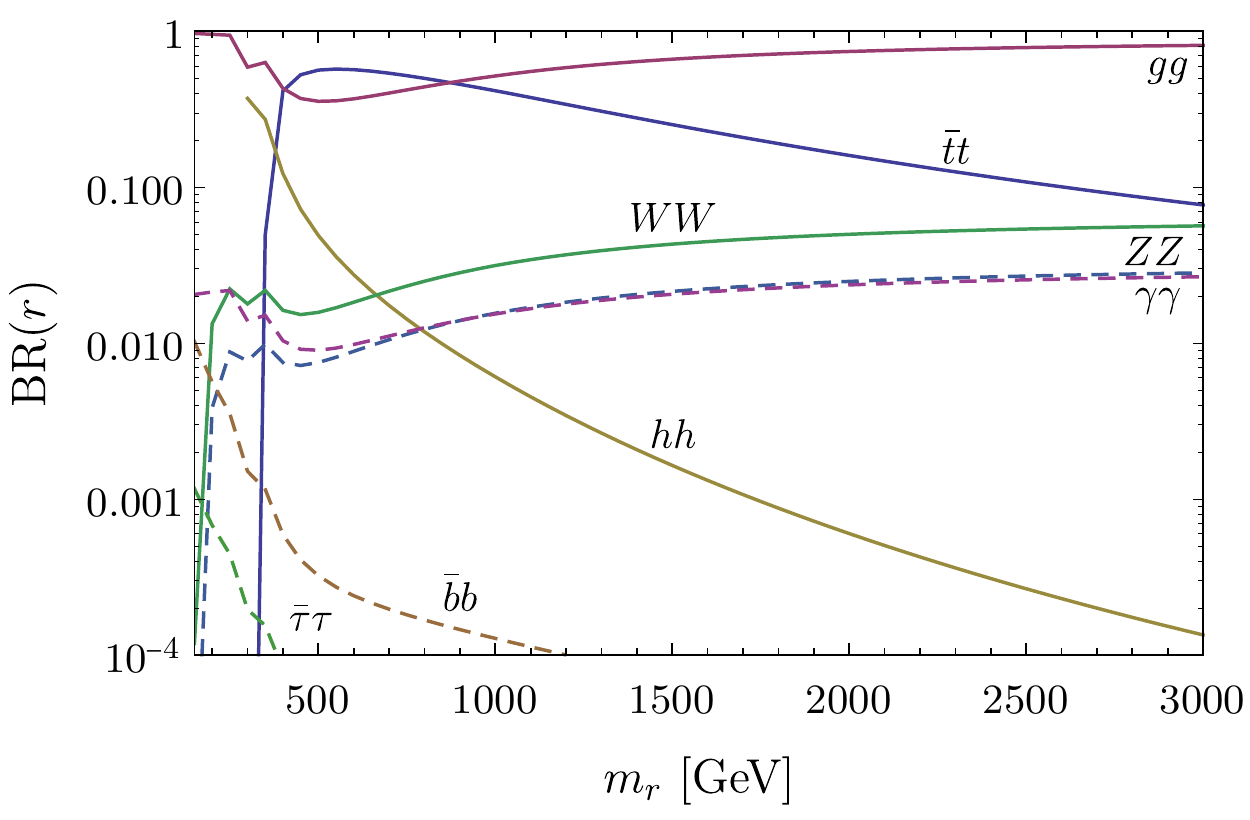}
 \end{minipage}
 \begin{minipage}[b]{0.5\textwidth}
  \centering
  \includegraphics[width=\textwidth]{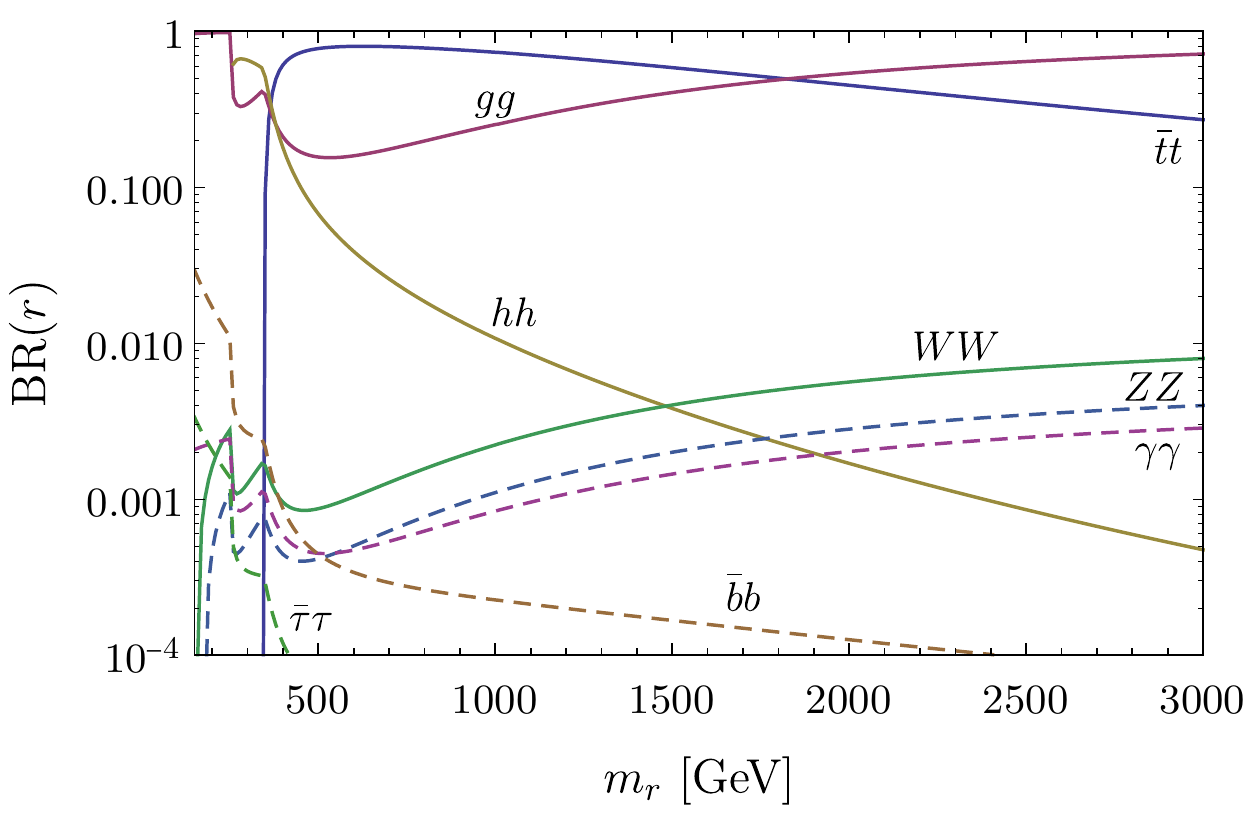}
 \end{minipage}
 \caption{\small The branching ratios of the radion to SM final states as a function of its mass and in the limit of zero mixing with the Higgs. The left panel is for $kL=7$ while the right panel is for $kL=35$.}
 \label{fig:br-mass}
\end{figure}

The branching ratios of the radion as a function of its mass are given in Figure~\ref{fig:br-mass}, in the limit where there is no mixing with the Higgs ($c_1=c_3=0$). The partial widths have been calculated using {\tt ehdecay}~\cite{Contino:2014aaa}, which incorporates higher order corrections for many of the final states\footnote{The expression for the $gg$ partial width implemented in {\tt ehdecay} is only valid for masses below 1~TeV. For larger radion masses we match onto the leading-order expression for the decay width. This results in a K-factor of 1.4, consistent with that obtained in Ref.~\cite{Bauer:2016lbe}}. We have also included by hand the terms proportional to $1/(kL)^2$ in the $WW/ZZ$ decay widths, since in our case these will generally provide the dominant contribution. We consider two values of the warp factor that span the range of allowed values; $kL \sim 7$ represents the lower limit obtained from flavour constraints, while the value $kL \sim 35$ addresses the electroweak-Planck hierarchy. As should be clear from Figure~\ref{fig:br-mass}, the branching ratios are dominated by the top and gluon couplings. On the other hand, the branching ratios to $\gamma\gamma/WW/ZZ$ significantly decrease as we move from $kL=7$ to $kL=35$ due to the $1/kL$ suppression of the couplings. The branching ratios remain qualitatively similar in the case of non-zero $c_1$ (while remaining in the alignment limit), with the exception of the branching ratio to $hh$. This decay can be significantly enhanced such that it competes with $gg$ and $\bar{t}t$ for large radion masses. Overall, Figure~\ref{fig:br-mass} implies that the channels expected to have meaningful sensitivity include the diphoton, ditop, dijet and possibly diHiggs final states. This should be contrasted with the usual radion paradigm with a brane localised Higgs, where decays to massive gauge bosons are relatively more important for collider searches. 

 In our analysis we include the most recent results from searches in the diphoton \cite{ATLAS:2016eeo, CMS:2016crm}, ditop \cite{CMS:2016ehh, ATLAS:2016-014}, dijet \cite{ CMS:2016ncz, ATLAS:2016lvi, CMS:2016wpz} and diboson \cite{ATLAS:2016kjy, ATLAS:2016cwq, ATLAS:2016bza, ATLAS:2016npe, CMS:2016pwo, CMS:2016knm, CMS:2016jpd, CMS:2016ilx} final states by both LHC collaborations. We should note that in the case of the ditop searches, the latest limits are only presented in terms of a spin-1 resonance. We will assume that the signal acceptance is unchanged in our case, which is a reasonable approximation\footnote{We do not expect interference effects to be important in our case~\cite{Carena:2016npr}.} based on the Run-I search from CMS~\cite{Chatrchyan:2013lca}. Regardless, it will turn out that this assumption has no effect on the current exclusion. The radion production cross section is calculated using the recent N$^3$LO results for a general scalar with both a Yukawa coupling to the top quark and an effective coupling to gluons~\cite{Anastasiou:2016hlm}.
 
\begin{figure}[t]
 \begin{minipage}[t]{0.5\textwidth}
  \centering
  \includegraphics[width=\textwidth]{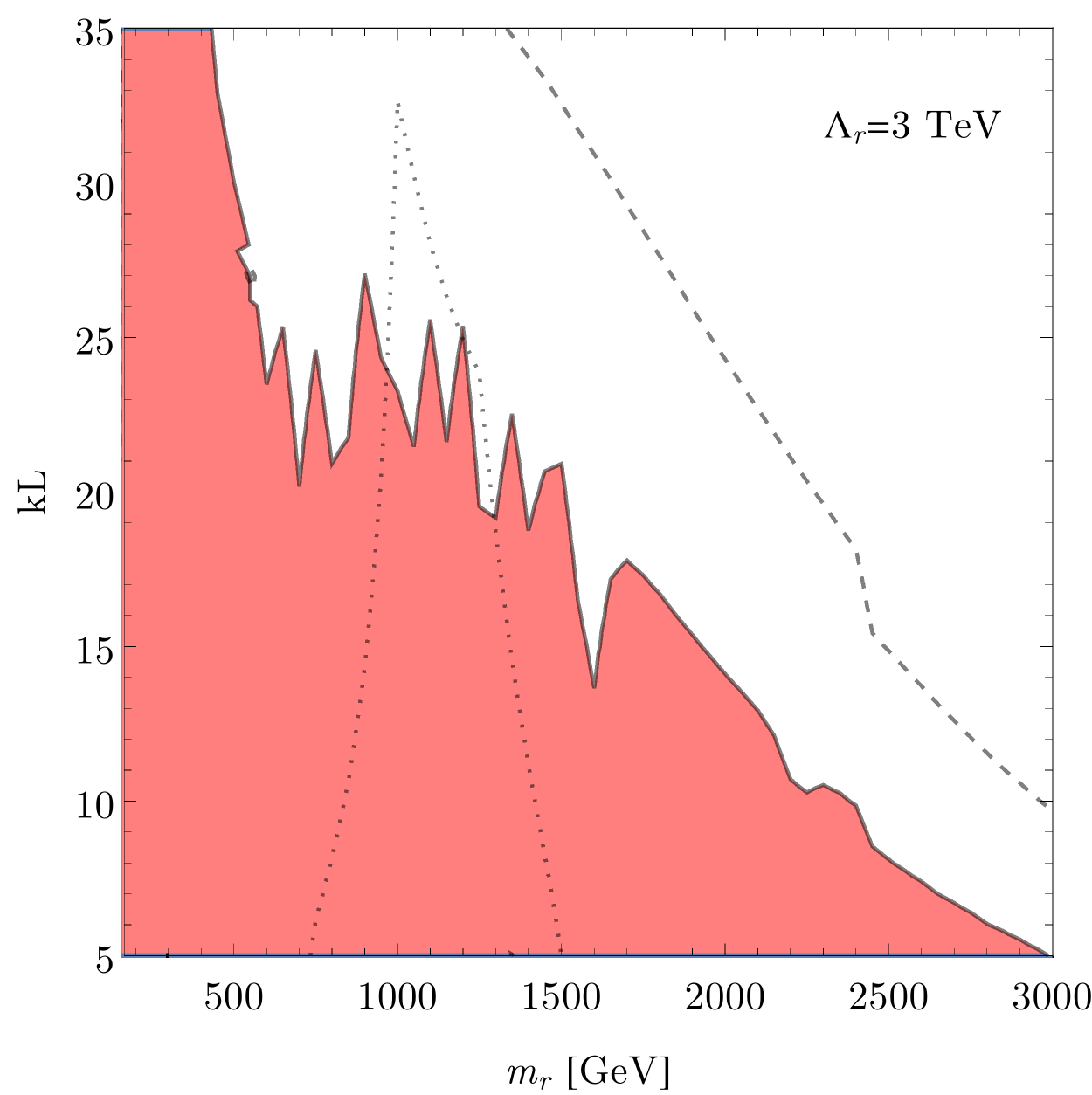}
 \end{minipage}
 \begin{minipage}[b]{0.5\textwidth}
  \centering
  \includegraphics[width=\textwidth]{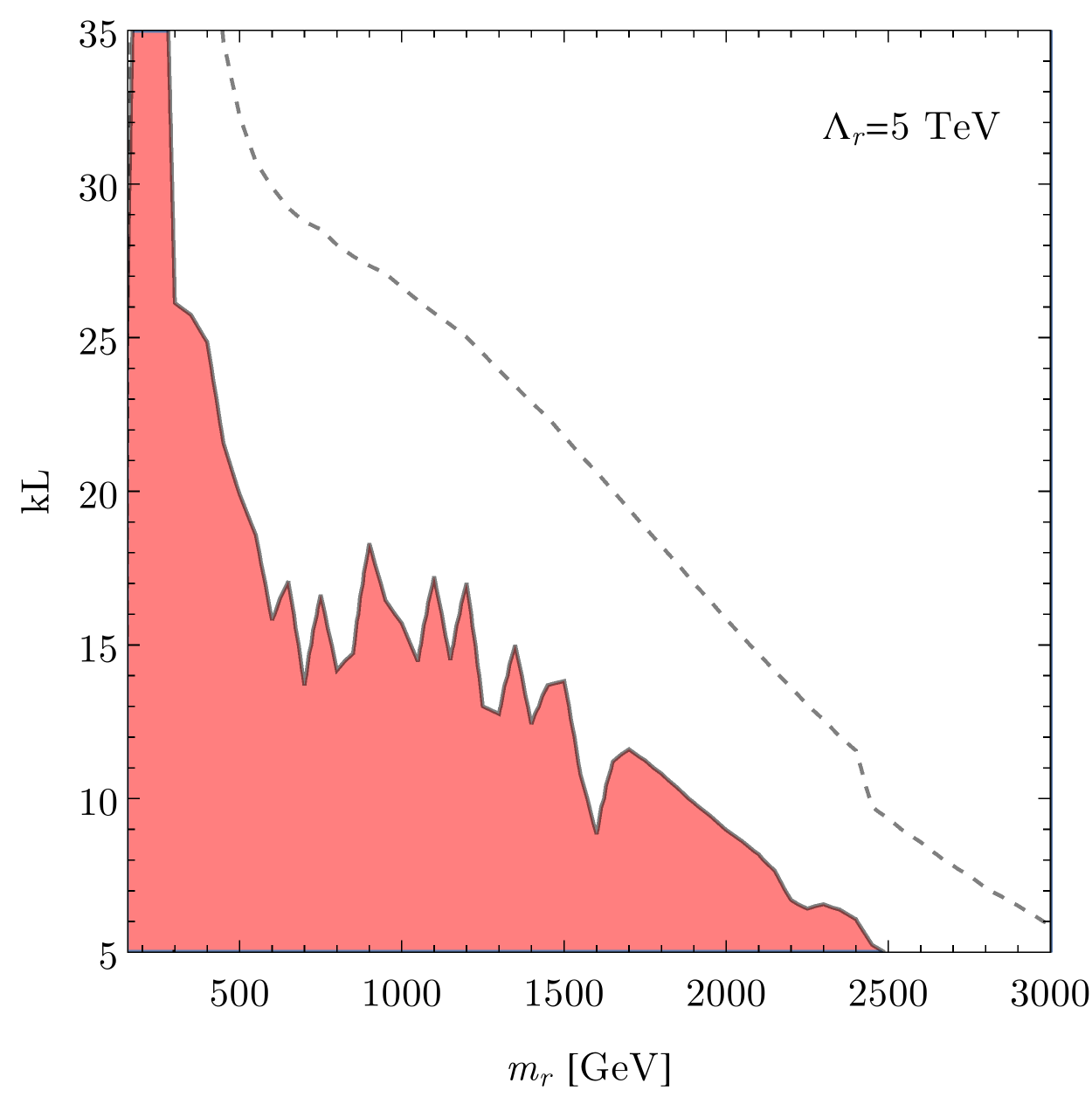}
 \end{minipage}
 \caption{\small Constraints on the parameter space in terms of the warp factor $(kL)$ and the radion mass $(m_r)$. The shaded region represents the present combined constraints from the LHC with $15\,\text{fb}^{-1}$. The dashed (dotted) contours represent the expected future reach with $300\,\text{fb}^{-1}$ from the diphoton (ditop) searches. The left (right) panels are for $\Lambda_r=3\,(5)\,$TeV and we have taken $c_1=c_3=0$.}
 \label{fig:mr_kL_const}
\end{figure}

\begin{figure}[t]
 \begin{minipage}[t]{0.5\textwidth}
  \centering
  \includegraphics[width=\textwidth]{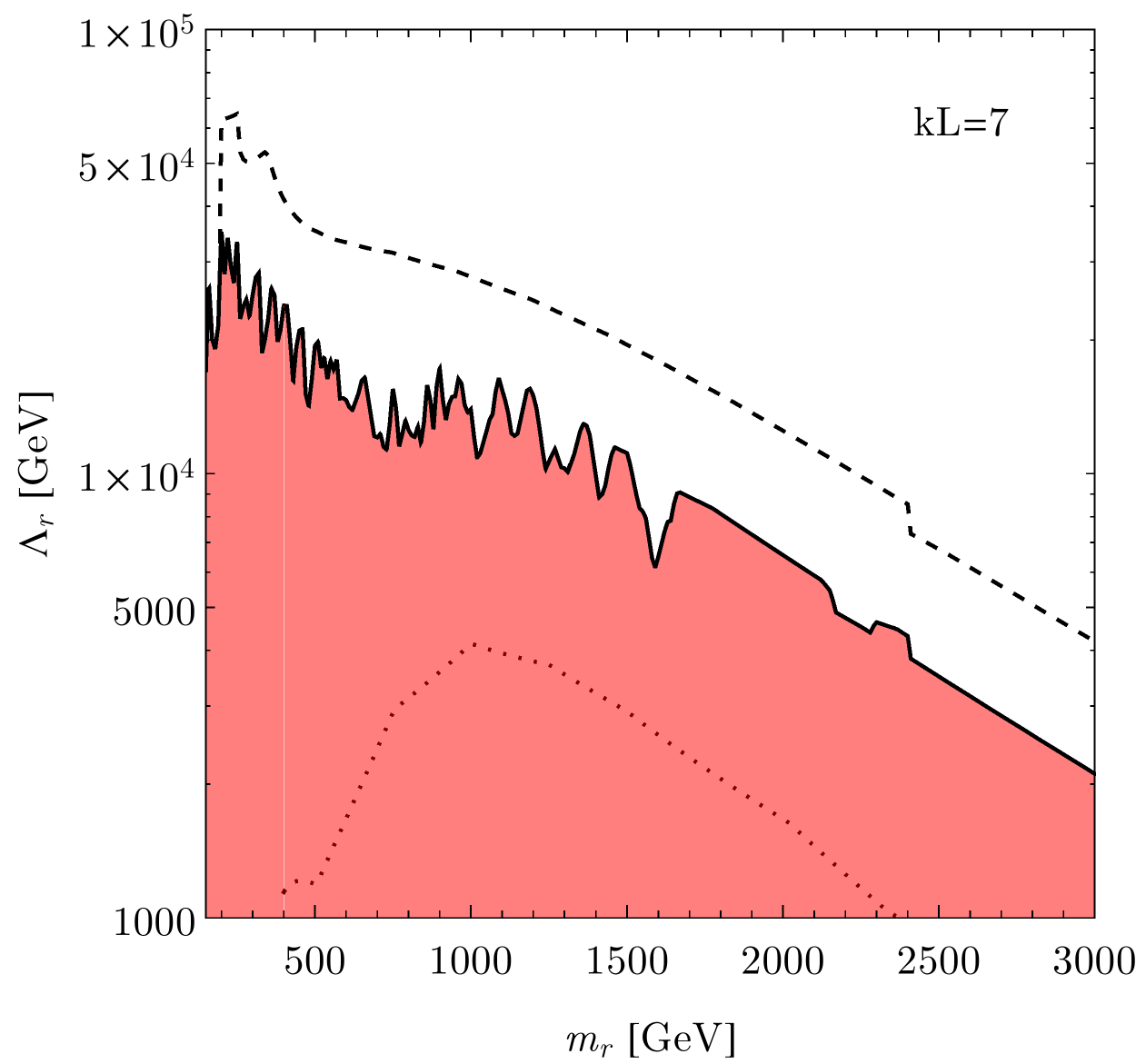}
 \end{minipage}
 \begin{minipage}[b]{0.5\textwidth}
  \centering
  \includegraphics[width=\textwidth]{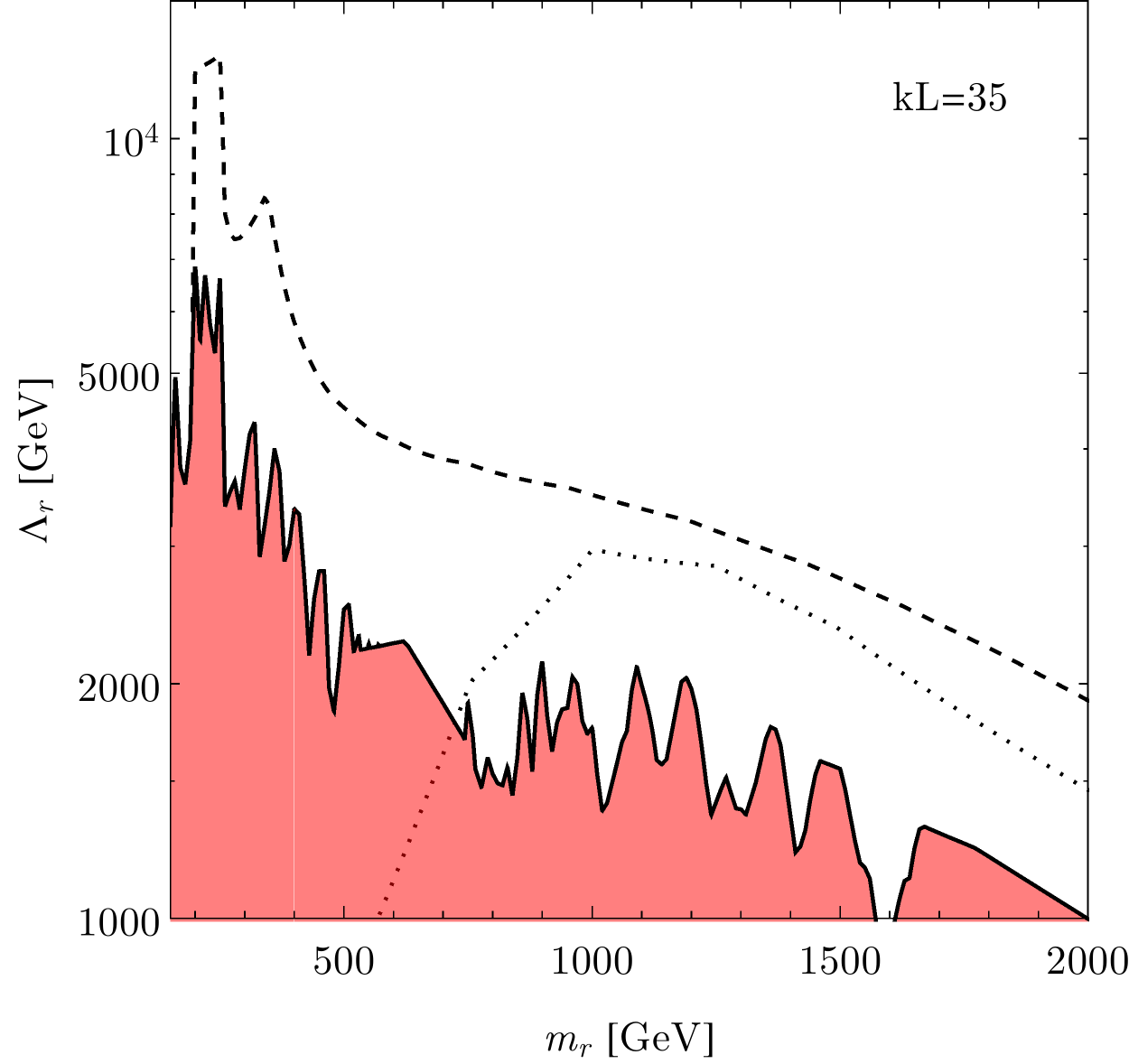}
 \end{minipage}
 \caption{\small Same as Figure~\ref{fig:mr_kL_const} but projected on to the parameter space of $\Lambda_r$ and $m_r$ for two extreme choices of the warp factor.}
 \label{fig:mr_lambda_const}
\end{figure}

The current combined constraint (red shaded region) on the warp factor $kL$ and radion mass $m_r$ parameter space with fixed values of $\Lambda_r = 3, 5$ TeV is shown in Figure~\ref{fig:mr_kL_const}. Similar constraints in the complimentary parameter space of $\Lambda_r$ and $m_r$ for fixed values of $kL= 7, 35$ are given in Figure~\ref{fig:mr_lambda_const}. While we have included the bounds from all potentially relevant final states, we find that the diphoton search provides the dominant constraint in almost all regions of the parameter space presented. The exception is a small region around $kL\sim 35$ and radion mass $m_r \sim 600$ GeV (see Figure~\ref{fig:mr_lambda_const}), where the 8~TeV ditop search still provides the strongest bound. The 13 TeV analyses in the $\bar{t}t$ final state currently only make use of the 2015 dataset ($\sim3\,\text{fb}^{-1}$) and so this channel may provide more competitive bounds in the future. A consequence of the dominance of the diphoton channel is that significantly weaker radion couplings (increased $\Lambda_r$) are already excluded in the case of smaller warp factors. This is due to the enhancement of the coupling to photons at small $kL$ and can be clearly seen in Figure~\ref{fig:mr_lambda_const}. If one considers the case $c_1\neq0$, searches in the $hh$ channel are also expected to provide a competitive constraint, especially for large values of $kL$. In this case the bounds from the diphoton search will be slightly weakened, particularly for larger radion masses.

In Figures~\ref{fig:mr_kL_const} \& \ref{fig:mr_lambda_const} we also show the projected reach of the ditop and diphoton channels (dotted and dashed lines respectively) at $\sqrt{s}=13\,$TeV with an integrated luminosity of $\mathcal{L}=300\,\text{fb}^{-1}$. These projections have been obtained by simply rescaling the current expected limits by $\sqrt{\mathcal{L}}$. Hence, they do not take into account any potential improvements in analysis techniques or future reductions in systematic uncertainties. Focusing on the left panel of Figure~\ref{fig:mr_kL_const} for $\Lambda_r=3$ TeV, we see that the ditop constraints start becoming relevant once that decay channel is kinematically available. The radion production cross section increases as $kL$ decreases, thus explaining the wider reach in radion mass for the ditop searches at smaller $kL$. Furthermore, notice that the slopes for the diphoton and ditop search are different in this figure. This can be understood given that as $m_r$ increases, $kL$ must decrease in order to maintain a sufficiently large radion production cross section. However, as $kL$ decreases the branching ratio to photons increases while the ditop branhcing ratio decreases, implying a milder slope for the diphoton search. This is just a consequence of the fact that the radion-diphoton and digluon couplings are (approximately) proportional to $1/(kL)$, while the coupling to top quarks is $kL$ independent. Moving to the right panel, we find that the ditop searches lose all sensitivity for larger values of $\Lambda_r$, regardless of the radion mass and warp factor. 

Focusing now on the right panel of Figure~\ref{fig:mr_lambda_const} for $kL=35$, we see a more similar fall off in sensitivity for the diphoton and ditop channels at large radion masses. This is a consequence of the fact that the radion couplings scale uniformly with $\Lambda_r$. The sensitivity of the ditop search also becomes significantly poorer as one moves towards the $\bar{t}t$ threshold, which leads to the peak in sensitivity around $m_r\sim1\,$TeV. For smaller values of $kL$, as shown in the left panel of Figure~\ref{fig:mr_lambda_const}, we see that the constraint from current diphoton searches already exceeds the projected sensitivity in the ditop channel. Even for the largest value of $kL$ considered, it is clear that diphoton searches will continue to dominate over the ditop channel with greater integrated luminosity, at least within the simple luminosity rescaling approximation we have performed.  This may change in the future with improved analysis techniques or reduced systematics in the ditop channel, esspecially given that this analysis is still relatively new at $\sqrt{s}=13\,$TeV.

\section{Conclusion}

In this work we have shown that the radion in the bulk Higgs warped extra dimensional scenario could have offered an explanation of the initially apparent diphoton excess at 750~GeV. The crucial observation is that the radion coupling to gauge bosons is suppressed compared to the more common brane Higgs models. This allows us to avoid the quite stringent constraints from searches for diboson resonances, provided that the mixing between the radion and the Higgs is small. 

We found that we could match the observed excess while avoiding all 8~TeV searches for a compositeness scale $\Lambda_r \gtrsim 2$~TeV and an extra dimension of size $kL \lesssim 20$. This relatively small $kL$ is necessary to enhance the radion coupling to diphotons. The dominant decay modes of the radion are to $t\bar{t}$ and $gg$, with the former providing both the most stringent bounds and the best prospects for probing this explanation for the diphoton excess. Mixing between the radion and the Higgs is already strongly constrained by $WW/ZZ$ resonance searches, which require being near the alignment limit $c_1=-c_3$. Finally, the radion branching ratio to $hh$ could be enhanced in certain regions of parameter space, potentially providing an additional channel to probe this model. 

Due to the final confirmation of the absence of an excess in the diphoton channel presented at ICHEP 2016 by both LHC collaborations, we extended our analysis to investigate the current and future constraints across the full radion parameter space. We find that diphoton searches currently provide the most stringent constraints on this scenario. A sizeable region of the parameter space has already been ruled out by the LHC Run-II results, in particular for smaller warp factors that can lead to large diphoton signals. In the case of non-zero $c_1$ (while remaining in the alignment limit), diHiggs searches may also provide an important, but more model-dependent constraint.  Looking to the future, we find that the diphoton channel is expected to continue to provide the most promising probe of these kinds of scenarios, with ditop and diHiggs searches providing a complementary approach.

\section*{Note added}

At the same time of posting this work on the arXiv, a similar work appeared~\cite{Ahmed:2015uqt} discussing the radion as a possible solution to the diphoton excess.

\section*{Acknowledgements}

ADM thanks Brando Bellazzini and Stephane Lavignac for useful discussions. TSR acknowledges the ISIRD Grant, IIT Kharagpur, India. This work was supported by IBS under the project code, IBS-R018-D1. This work has been supported in part by the European Research Council (ERC) Advanced Grant Higgs@LHC. This work was supported by the Australian Research Council.

\end{document}